\documentclass[pra,aps,twocolumn,showpacs,showkeys,amsmath,amssymb,reprint,floatfix]{revtex4-1}

\usepackage{graphicx}
\usepackage{color}
\usepackage{bm}
\usepackage{hyperref}

\newcommand{\Hch}{\ensuremath{H_{\textrm{ch}}}}

\begin{document}

\title{Chiral bound states in the continuum}

\author{Jordi Mur-Petit}
\email{jordi.mur@csic.es}
\affiliation{Instituto de Estructura de la Materia, IEM-CSIC, Serrano 123, Madrid 28006, Spain}

\author{Rafael A. Molina}
\affiliation{Instituto de Estructura de la Materia, IEM-CSIC, Serrano 123, Madrid 28006, Spain}

\begin{abstract}
We present a distinct mechanism for the formation of Bound states In the
Continuum (BICs). In chiral quantum systems there appear zero-energy states in which the wave function has finite amplitude only in one of the subsystems defined by the chiral symmetry. When the system is coupled to leads with a continuum energy band, part of these states remain bound.
We derive some algebraic rules for the number of these states depending on the dimensionality and rank of the total Hamiltonian.
We examine the transport properties of such systems including the appearance of Fano resonances in some limiting cases.
Finally, we discuss experimental setups based on microwave dielectric resonators and atoms in optical lattices where these predictions can be tested.
\end{abstract}


\maketitle

\section{Introduction}\label{sec:intro}

Interference effects account for some of the surprises found in quantum physics. In the first years of the study of quantum mechanics, von Neumann and Wigner showed that some particular cases of spatially oscillating attractive potentials support so-called bound states in the continuum (BICs): square-integrable solutions of the time-independent Schr\"odinger equation with eigenenergies above the potential threshold~\cite{vonNeumannWigner29}.
BICs can be formed above the barrier because there appears a destructive interference between the different partial wave amplitudes that causes the wave function to vanish at large distances~\cite{Weber94}.
Given the peculiar long-range oscillating behavior of the potentials constructed by Wigner and von Neumann, for a long time BICs were considered as nothing more than a mathematical curiosity. More important, as these states are true eigenstates of the Hamiltonian---in contrast with Fano-Feshbach resonances~\cite{Kohler06RMP}, they are not related to near-threshold states of an approximate Hamiltonian~\cite{Friedrich85b}---they are embedded in the continuum but not coupled to it, so it appeared unclear how one could actually probe them.


However, in 1985 Friedrich and Wintgen proved that two coupled resonant states connected to a continuum could lead to a BIC if one could drive them into degeneracy, e.g., using an external field~\cite{Friedrich85b}. A contemporary experiment indeed observed such a phenomenon as a reduced auto-ionization rate of a Rydberg state in barium~\cite{Neukammer85}.
More recently, following the proposals by Herrick~\cite{Herrick1976} and Stillinger~\cite{Stillinger1976} based on the analogy between the time-independent Schr\"odinger equation and the wave equation, technological advances have allowed the observation of bound states above the continuum threshold in semiconductor quantum wells~\cite{Capasso92} and photonic systems~\cite{Plotnik11}.
These observations have triggered  a revival in the study of BICs due to their potential applications for quantum transport and classical or quantum information devices.

From a theoretical perspective, the occurrence of BICs has been investigated in different systems, including mesoscopic structures~\cite{Schult90,Rotter05,Ladron06}, the hydrogen atom in a uniform magnetic field~\cite{Friedrich85}, and photonics~\cite{Marinica08,Bulgakov08,Prodanovic09}.
Several mechanisms have been identified for the formation of BICs beyond the original work by von Neumann and Wigner.
For example, BICs can appear when a discrete eigenstate and the continuum spectrum have different parities~\cite{Texier02,Voo06}. BICs can also occur due to an exact destructive interference process occurring for specific values of the parameters in models of quantum dots~\cite{Miyamoto05,Sadreev06,Bulgakov07}.

In this work we identify a very general class of systems---systems presenting chiral symmetry---that can support BICs when connected to a continuum.
After discussing the principles underlying the existence of these states, we show general rules for the number of BICs that can be present in the middle of the energy band of the system. Chiral systems include a very general class of hopping Hamiltonians within a bipartite lattice, and we show some examples of such systems using models of quantum dots defined in a bipartite lattice and connected to left and right leads.
We calculate the conductance in a two-terminal setup and the density of states as a function of the energy showing the effects of the BICs in the transport properties of the system for both square and honeycomb lattices (Sect.~\ref{sec:transp}).
In Sect.~\ref{sec:disc} we discuss how the theory that we present allows to revisit in a more general perspective earlier results on the conduction of quantum-dot structures and the appearance of Fano resonances in periodic nanoparticle arrays.
Finally, in Sect.~\ref{sec:expt} we discuss how these states could be observed in different experimental setups, from quantum-dot arrays, similar to Refs.~\cite{Capasso92,Plotnik11}, to photonic resonator lattices~\cite{Bellec13}, or cold atoms in optical lattices~\cite{Brantut2012,Brantut2013,Krinner2014}.

\section{Chiral Hamiltonians and chiral eigenstates}\label{sec:chiral}

\begin{figure}[tb]
 \centering
  \includegraphics[width=\linewidth]{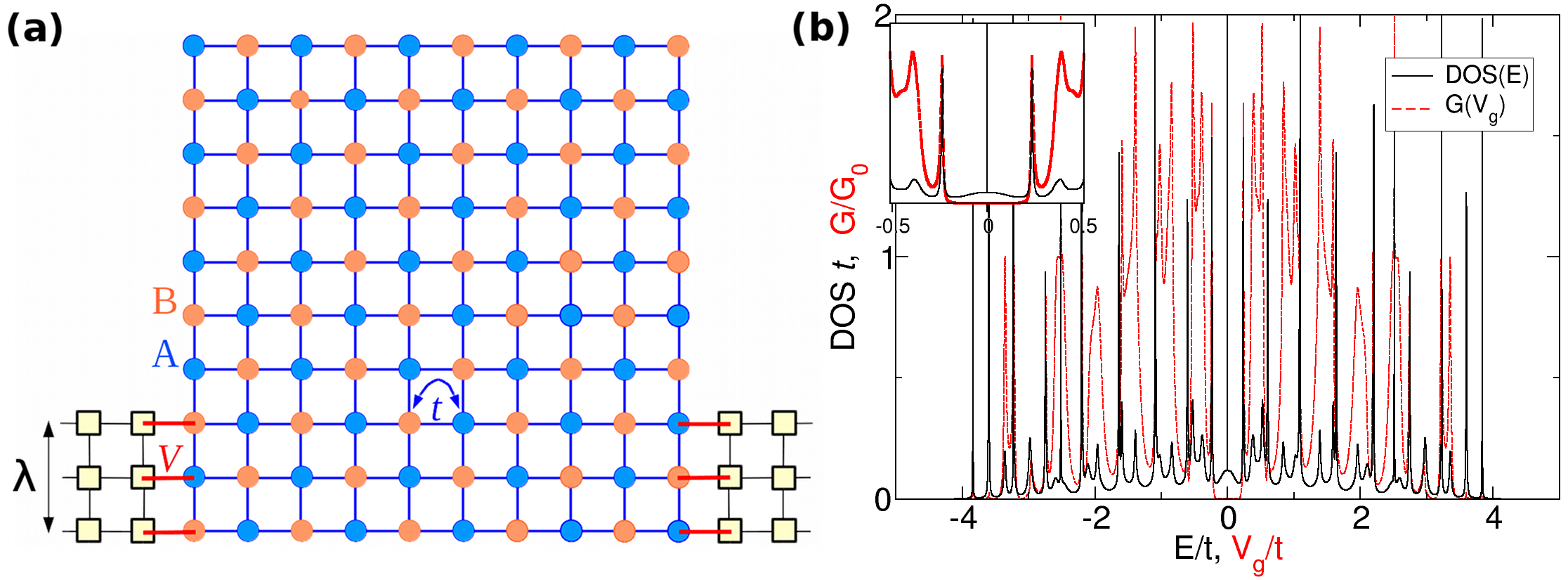}
  \caption{\label{fig:square_lattice}
     (color online)
     \textbf{(a)} Closed chiral system (circles) with hopping amplitude $t$, linked with coupling strength $V$ to quasi-1D leads of size $\lambda$ (squares). On-site energies for sites in the closed system are displaced uniformly by a gate potential $V_g$ and the conductance is calculated as a function of $V_g$.
     \textbf{(b)} Conductance $G(V_g)$ (red or gray line) and density of states $\rho(E)$ (black) for the system with $\lambda=3$ leads in (a).
     The inset is a close up to the region around $E=0$.
  }
\end{figure}

We consider a generic chiral system connected to left and right leads,
\begin{equation}\label{H:General}
 H=\Hch + H^{L}_\textrm{lead}+H^{R}_\textrm{lead}+H_\textrm{ch-lead} \:,
\end{equation}
where $\Hch$ denotes the Hamiltonian of the chiral system, $H^{L(R)}_\textrm{lead}$ is the Hamiltonian of the left (right) lead, and $H_\textrm{ch-lead}$ contains the coupling terms between the system and the leads.
The Hamiltonian of a general chiral system can be written
\begin{align}\label{H:chiral}
 \Hch &= \left( \begin{array}{cc} 0 & C \\ C^T & 0 \end{array} \right) \:.
\end{align}
Here, $C^T$ stands for the matrix transpose of $C$.
The system becomes open by connecting it to leads carrying incoming and outgoing scattering states. We assume quasi-1D leads, i.e., their eigenstates are momentum, $k$, states supported on a finite number of transverse channels,
$\lambda_\alpha$ ($\alpha=L,R)$, with annihilation (creation) operators $a^{(\dagger)}_{k^{(n)}_\alpha}$, 
\begin{equation}\label{H:lead}
 H^{\alpha}_\textrm{lead}
 =
 \sum_{n=1}^{\lambda_{\alpha}}
 \sum_{k^{(n)}_{\alpha}}
 \epsilon_{k^{(n)}_{\alpha}} a^{\dagger}_{k^{(n)}_{\alpha}} 
a^{\phantom{\dagger}}_{k^{(n)}_{\alpha}},
 \quad
 \alpha=L,R \:.
\end{equation}
We assume that the coupling between the lead states and the system happens directly between a finite number, $\lambda$, of states in the two chiral regions ($A,B$) and the momentum states of the leads [see Fig.~\ref{fig:square_lattice}a],
\begin{align}\label{H:ch-lead}
 H_\textrm{ch-lead}
 &=
 \sum_{i\in A,B} 
 \sum_{\alpha=L,R} \sum_{n=1}^{\lambda_{\alpha}} 
 \sum_{k^{(n)}_{\alpha}} V_{i; k^{(n)}_{\alpha}} 
 c^{\dagger}_{i} a^{\phantom{\dagger}}_{k^{(n)}_{\alpha}}
+ \textrm{H.~c.}
\end{align}
Here $i$ runs over the system Hamiltonian $\Hch$ eigenstates,
and $V_{i; k^{(n)}_{\alpha}}$ is the coupling strength between system state $i$ and the state with momentum $k$ on transverse channel $n$ of lead $\alpha$.
A typical example of a system described by this model is a quantum system characterized by a nearest-neighbor tight-binding Hamiltonian on a bipartite lattice connected to left and right leads through a finite number of sites on the boundary, see Fig.~\ref{fig:square_lattice}a.

\subsection{Chiral eigenstates in closed systems}\label{ssec:closed}

To identify the BICs, we first find the eigenstates of the closed system.
We start by solving the time-independent Schr\"odinger equation for the closed system introducing a two-component wavefunction to represent the distribution on the two chiral subsystems,
\begin{eqnarray}\label{eq:closed}
 \Hch\Psi_\textrm{ch}=
 \left( \begin{array}{cc} 0 & C \\ C^T & 0 \end{array} \right)
 \left( \begin{array}{c} \Psi_A \\ \Psi_B \end{array} \right)
= E \left( \begin{array}{c} \Psi_A \\ \Psi_B \end{array} \right),
\end{eqnarray}
being $A$ and $B$ the two chiral subsystems (the two sublattices of the bipartite lattice in our previous example).
It is easy to show, from basic algebraic considerations, that the spectrum of the system is symmetric under the operation $E \rightarrow -E$:
the wavefunction of the eigenstate with energy $E$ is related to the wavefunction at $-E$ by a change of sign of the components in one of the subspaces.

Let us focus our attention on the states at the band center, $E=0$, of the closed system.
Finding the eigenstates at zero energy is equivalent to solving a homogeneous system of linear equations.
The degeneracy of the subspace of states with $E=0$ is the nullity, ${\cal N}_\textrm{ch}$, of $\Hch$.
Now, the chiral structure of the Hamiltonian allows us to relate the number of zero-energy eigenstates with the properties of the matrix $C$~\cite{Inui94,Victor13}.
This follows from the rank-nullity theorem from linear algebra, which states that the nullity plus the rank of a matrix is equal to the dimension of the matrix~\cite{ranknullity}.
Hence a basis on the subspace of solutions of the homogeneous system of equations can be found, such that each basis state is non-vanishing on only one of the sublattices.
More precisely, the dimension of each sublattice, $n_{A,B}$, minus the rank of the matrix $C$ (the same as that of $C^T$) equals the dimension of the zero-energy subspace on that sublattice, ${\cal N}_{A,B}$.
The total nullity of $\Hch$ is then ${\cal N}_\textrm{ch}={\cal N}_A+{\cal N}_B$.

The minimum value of ${\cal N}_\textrm{ch}$ is, therefore, equal to the difference between the dimensions of the $A$ and $B$ subspaces ${\cal N}_\textrm{ch} \geq |n_{A} - n_{B}|$.
When the inequality is saturated, all zero-energy eigenstates have components different from zero only on the sublattice with the largest number of sites. 
In general, in a disordered chiral system the inequality is saturated. Then, if the number of sites in both sublattices is the same, there are no zero-energy eigenstates~\cite{Inui94}. On the other hand, if the number of sites is bigger in one of the sublattices, for example $A$, then ${\cal N}_\textrm{ch} ={\cal N}_A = n_{A} - n_{B}$, while ${\cal N}_B=0$.
By contrast, in the case of ordered lattices, the symmetries within $C$ make it possible that ${\cal N}_A+{\cal N}_B \gg |n_{A} - n_{B}|$. In this case, the total nullity can be large even for $n_{A}=n_{B}$, depending on the boundary conditions of the system~\cite{Victor13}.


\subsection{Open systems: Chiral BICs}\label{ssec:open}

We proceed now to present the main result of this paper: 
Once we connect the leads to the system, part of the zero-energy subspace found in Sect.~\ref{ssec:closed} may remain bound even when $E=0$ lies within the continuum of states on the leads. When this happens, these states form a special kind of BICs that we call \textit{chiral BICs}.

The wavefunction of an arbitrary state of the open system can be written as
\begin{equation} \label{eq:decomposition}
 \Psi(t)=\sum_r w_r(t) \Psi_r \:, 
\end{equation}
where $\Psi_r$ are the eigenstates of the system together with the leads, which form a basis of the Hilbert space corresponding to $H$. They satisfy $H\Psi_r = E_r\Psi_r$, 
with $r$ the quantum number identifying state $\Psi_r$; it runs over all states of the open system.
Using Eq.~\eqref{eq:decomposition}, the time-dependent Schr\"odinger equation of an arbitrary state can be expressed as
\begin{equation}\label{eq:dwrdt}
 i \hbar \frac{d w_r(t)}{dt}
 = \sum_s H_{rs} w_s (t) \:, 
\end{equation}
where $H_{rs}=\langle \Psi_r|H| \Psi_s\rangle$ is the Hamiltonian matrix element between eigenstates $r$ and $s$ of the open system.

Let us ignore inelastic processes and, recalling that we are looking for states of zero energy, restrict ourselves to lead states with energy $E=0$, i.e., $k=0$.
Then,
\begin{equation}\label{eq:wr}
 i \hbar \frac{d w_r(t)}{dt}
 = \sum_s \left[
      \left(\Hch\right)_{rs}
     + \sum_{\alpha=L,R} \sum_{n=1}^{\lambda_{\alpha}} V_{s;(k=0)^{(n)}_\alpha}
   \right] w_s \:.
\end{equation}
Taking into account only the $k=0$ subspace in the leads, the index $r$ is now restricted to $r=1,\ldots,L+\lambda_L+\lambda_R$,
where $L$ is the size of the (closed) system, and we recall that $\lambda_{\alpha}$ stands for the number of transverse channels of the lead $\alpha=L,R$ attached to the system.
The set of equations~\eqref{eq:wr} constitutes a homogeneous system of $L+\lambda_L+\lambda_R$ linear equations for the coefficients $w_r$.

We are interested in finding stationary solutions of the open-system Schr\"odinger equation that do not populate states in the right or left leads,
i.e., we seek eigenstates of $H$ whose wavefunction can be expanded as in Eq.~\eqref{eq:decomposition} with $r$ running over states that are localized only on the (closed) system.
As the closed system cannot support more than $L$ linearly independent states, we take these to be the basis states $\Psi_r$ with $r=1,\ldots, L$
and impose $w_r=0$ for $r>L$ in Eq.~\eqref{eq:decomposition}.
Recalling that we are interested in zero-energy eigenstates, we have
$i\hbar d\Psi/dt=H\Psi=0$ which, through Eq.~\eqref{eq:decomposition}, implies $dw_r/dt=0$.
Substituting this into Eq.~\eqref{eq:wr} results in:
\begin{align} \label{eq:open}
 \sum_{s=1}^L \left[ \left(\Hch\right)_{rs}
    + \sum_{\alpha=L,R} \sum_{n=1}^{\lambda_{\alpha}} V_{s;(k=0)^{(n)}_\alpha}
 \right] w_s = 0 \:,
 \nonumber
 \\ (r=1, \ldots, L+\lambda_L+\lambda_R) \:.
\end{align}
This is a set of $L+\lambda_L+\lambda_R$ equations for 
$L$ unknowns (the coefficients $w_r$ with $r=1, \ldots, L$), i.e., Eq.~\eqref{eq:open} is an overdetermined homogeneous system of equations.
The number of its fundamental solutions depends on the rank of the open-system Hamiltonian matrix, Eq.~\eqref{H:General}.

From elementary algebraic considerations, it is easy to see that, for a lattice system with only nearest-neighbor hopping attached to quasi-1D leads, the rank of $H$ will generally be increased from that of $\Hch$ by the number of sites of each sublattice connected to the leads, reducing the number of fundamental solutions of Eq.~\eqref{eq:open} by the same amount. Exceptions to this rule may occur in cases with specific degeneracies between the couplings to the leads and parameters inside $\Hch$; in such situations, a full understanding of the properties of the system demands a case-by-case analysis.

Leaving aside such degenerate cases, we can proceed by recalling from the discussion in Sect.~\ref{ssec:closed} that the number of zero-energy eigenstates ${\cal N}_\textrm{ch}$ of the closed system is divided between states in subsystem $A$ (${\cal N}_A$) and states in subsystem $B$ (${\cal N}_B$).
It follows that, on connecting the system to the leads, each of ${\cal N}_{A,B}$ will be reduced by the number of sites in the corresponding sublattice connected to the leads.
A case of particular interest occurs if the leads are connected to sites on only one of the sublattices. Then, all the zero-energy states living on the other sublattice will remain bound solutions of the open-system Schr\"odinger equation. These remaining eigenstates of $H$ are true BICs: they do not contribute to the transport of current and do not decay~\cite{Victor13}.


\section{Chiral BICs and transport properties of bipartite systems}\label{sec:transp}

From the previous considerations, we conclude that chiral systems will generally present BICs at zero energy. As we open the system, for example for two-terminal conductance measurements, the zero-energy subspace will divide itself between a part connected to the leads and a part disconnected from the outside. The former will be displaced in energy due to the coupling to the leads, and so it will not affect the measurements on the band center, where signatures of the latter can be looked for. Thus, a generic characteristic of transport through chiral systems (with openings connected to a small number of sites compared to the total number of sites in the system) is that they present a dip in the conductance as a function of the energy (or, equivalently, as a function of some gate potential) in the middle of the band.

\subsection{Clean square lattice}\label{ssec:results}

To illustrate the implications of the discussed properties of chiral systems, we study a system consisting of a number of discrete points arranged on a square lattice with a square shape of $10 \times 10$ sites as shown in Fig.~\ref{fig:square_lattice}a.
The system is described by a hopping Hamiltonian,
\begin{equation}\label{eq:hubbard-ham}
 \Hch = -t \sum_{\langle i,j \rangle}
   \left( c_i^{\dagger} c_j + \textrm{H.c.} \right)  \:,
\end{equation}
where the sum is over nearest neighbors (which lie on different sublattices), and $t$ is the hopping amplitude.
This system can stand as a simple model for an array of quantum dots~\cite{Capasso92,Plotnik11} or an array of dielectric resonators in a microwave cavity~\cite{Bellec13}.
If we calculate the rank of the matrix or solve the closed-system Schr\"odinger equation by exact diagonalization, we obtain $10$ zero-energy states. We can then connect the system to leads in many different ways. As an example, we first choose to join identical leads ($\lambda_L=\lambda_R \equiv \lambda$) to the three bottom sites on the left and right sides of the square as shown in Fig.~\ref{fig:square_lattice}a. We calculate the conductance using the Landauer-B\"uttiker formulation with the usual prescription for calculating the transmission from the Green's function in tight-binding lattices including a gate potential, $V_g$, that modifies uniformly the potential energy in the system~\cite{Datta_book}. In our calculations below, we set the value $t=1$, and model the coupling to the leads by hopping terms from sites in the system to sites in the leads with amplitudes $V=1$.

In Fig.~\ref{fig:square_lattice}b we show a plot of the conductance, $G(V_g)$, and the density of states (DOS) of such a quantum dot connected to leads.
(In this and the following figures, energies are given in units of the hopping energy $t$, while the density of states is plotted in units of $1/t$ and the conductance is displayed in units of the quantum of conductance $G_0=e^2/h$.)
As expected, the figure is symmetric under a reflection through $E=0$. Each of the peaks in the density of states can be correlated with a corresponding peak in the conductance except precisely at $E=0$, the position of the chiral BICs. The inset of Fig.~\ref{fig:square_lattice}b shows a zoom at the region around $E=0$ revealing the narrow peak in the DOS at $E=0$ while the conductance remains negligible.

\begin{figure}[tb]
 \centering
 \includegraphics[width=\linewidth]{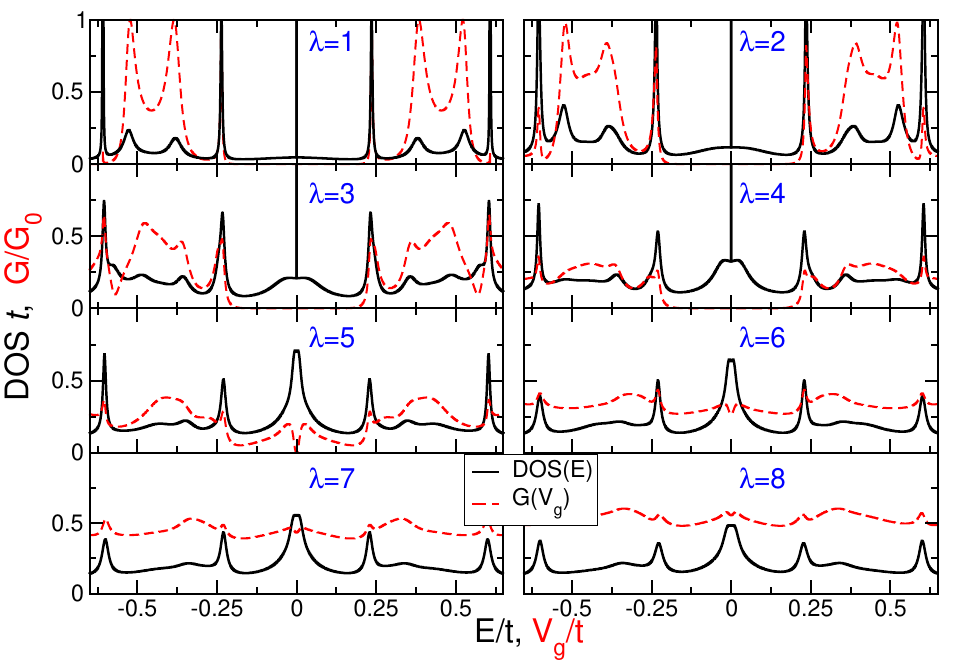}
 \caption{\label{fig:dosG}
  (color online)
  Density of states, DOS($E$), (black solid lines) and conductance, $G(V_g)$, (red or gray dashed) for a $10\times 10$ system connected to leads of different sizes, $\lambda=1,\ldots,8$, as indicated.
  Note the broad non-conducting band centered around $E=0$ for $\lambda \leq 4$, and the Fano antiresonance dip on $G(V_g)$ for $\lambda=5$.}
\end{figure}

The evolution of the conductance and DOS curves
as we increase the number of channels in the leads is shown in Fig.~\ref{fig:dosG}, starting with only one-channel leads and up to $\lambda=8$ leads, for the same $10 \times 10$ system.
As the number of degenerate zero-energy eigenstates for the closed square is ${\cal N}_\textrm{ch}=10$ (5 on each sublattice), we expect to have BICs for up to $\lambda=4$ channels, while for $\lambda \geq 5$ all BICs should disappear. This is exactly what we observe in the figures:
the conductance remains close to zero over a wide gate-voltage range around $V_g=0$, indicating non-conducting behavior, for $\lambda=1\ldots4$, while the DOS features a narrow peak just at $E=0$---this points to a number of zero-energy eigenstates that do not couple to the leads, \textit{i.e.}, BICs. In addition, for increasing $\lambda$ we see how the DOS acquires a broader peak around the band center, which however does not lead to an increase in conductance in the equivalent range of gate voltages.

For $\lambda=5$ channels, the DOS no longer has a narrow peak and the conductance shows a shape reminding of a Fano antiresonance~\cite{Miros10}. This characteristic can be understood as due to the last remaining BIC becoming coupled to the continuum of states on the leads. For a larger number of outcoupling channels, the conductance at $V_g=0$ ceases to vanish, i.e., the system becomes a conductor, and the DOS peak becomes progressively broader and lower.

\subsection{Disordered systems}\label{ssec:disorder}

\begin{figure}[tb]
 \centering
 \includegraphics[width=\linewidth]{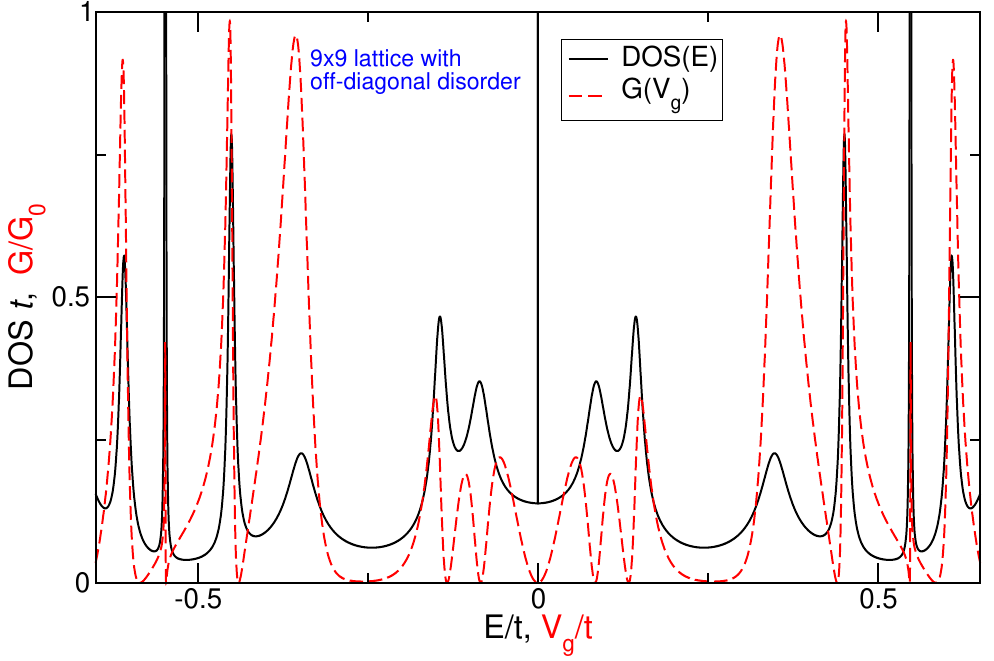}
 \caption{\label{fig:disorder}
 (color online)
 Conductance (red or gray dashed line) and DOS (black solid) for a $9 \times 9$ square lattice with two quasi-1D leads connected to the smaller sublattice and with off-diagonal disorder, $t\in[0.5,1.5]$.}
\end{figure}

As mentioned earlier, in the case where one of the sublattices supports a larger
number of states than the other sublattice, the difference between sublattice dimensions equals the minimum number of zero-energy eigenstates. In that case, even for disordered systems we can have BICs if the leads are connected only to the sublattice with a smaller number of sites. To check this prediction, we have calculated the conductance of a $9 \times 9$ lattice with leads connected to the second lowest sites on the left and right sides of the square, which lie in the smaller sublattice.
We use a particular realization of a disordered lattice Hamiltonian with hopping matrix elements distributed uniformly in the range $[0.5,1.5]$. The conductance as a function of the gate potential and the density of states as a function of the energy are shown in Fig.~\ref{fig:disorder} on top of each other as in the previous figures. The DOS diverges at $E=0$ while the conductance vanishes at $V_g=0$ marking the presence of a BIC in the middle of the band. As the disorder is only off-diagonal, it conserves the chiral symmetry, and the curves are symmetric under $E \rightarrow -E$.

\subsection{Chiral BICs in the honeycomb lattice}\label{ssec:honeycomb}

\begin{figure}[tb]
 \centering
 \includegraphics[width=\linewidth]{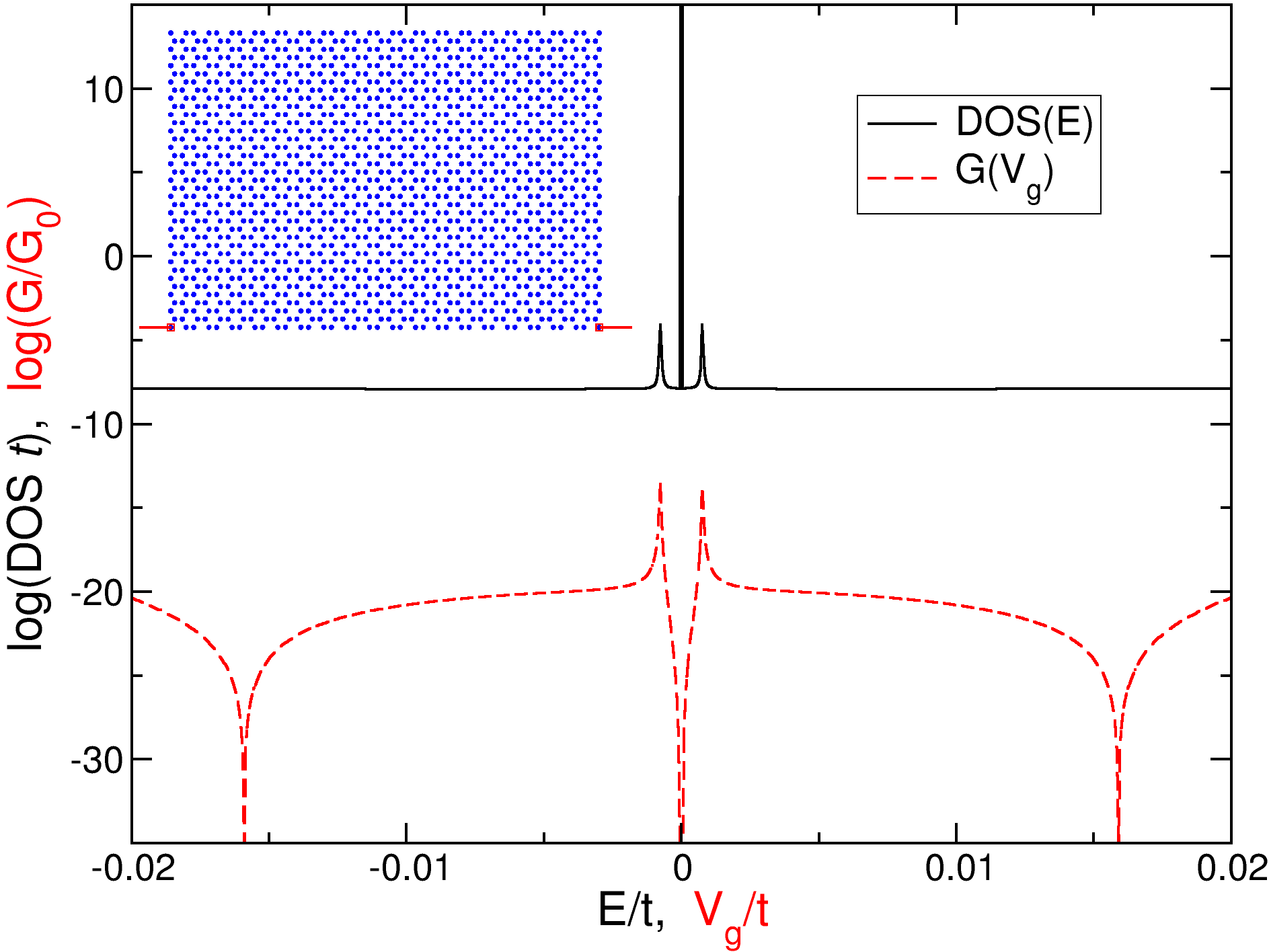}
 \caption{\label{fig:graphene}
 (color online)
 Conductance (red or gray dashed line) and DOS (black solid) for a $38 \times 38$ square honeycomb-lattice flake with two leads connected to sites on opposite zig-zag edges, as marked in the inset (top left). Note the semi-logarithmic scale.}
\end{figure}

The results presented are applicable to any bipartite lattice. In particular, zero-energy eigenstates on the honeycomb lattice correspond to the edge states that appear in the case of zig-zag edges~\cite{Fujita96}. In that case, the edge states can become BICs if we join a zig-zag graphene nanoribbon with properly placed leads. We present an example in Fig.~\ref{fig:graphene} where we show results for a $38\times 38$ square graphene flake connected to leads by only one carbon atom in opposite edges (see inset). Although, this situation is probably unrealistic for actual graphene it is certainly feasible for artificial honeycomb lattices~\cite{Bellec13}.

For such a flake, there are no zero-energy eigenstates on the (top and bottom) armchair edges, that are formed by sites belonging to both sublattices; this is in agreement with our findings above that zero-energy eigenstates are localized on one of the sublattices. On the other hand, there will be eigenstates at $E=0$ localized on the (left and right) zig-zag edges; these are indeed edge states living only on one of the sublattices of the honeycomb lattice. There can be current along each of these edge states; however, there will be no current between leads connected to opposite edges.

These expectations based on our arguments in the previous paragraphs are confirmed by 
numerical calculations. We have calculated the density of states and the conductance when two leads are connected to single sites on opposite zig-zag edges of a $38\times 38$ flake; our results are shown in Fig.~\ref{fig:graphene} using a semi-logarithmic scale as the conductance close to the band center is quite small. At the band center, the DOS diverges, marking the presence of the edge states, while there is a zero of the conductance, thus confirming that the edge states behave as BICs with this configuration of the leads.

The two small conductance peaks at $0<|V_g|\ll t$ are related to resonances with very small probability density in the sites connected to the leads. Other transmission zeroes marked by dips in the conductance are unrelated to BICs but due to consecutive resonances with the same parity in the one-channel case~\cite{LevyYeyati00,Silva02,Molina12,Molina13}.

The interplay between the geometry of the honeycomb flake, the number of zig-zag edges, and the position of the sites connected to the leads can give rise to very different transport properties close to the band center. A detailed analysis should be quite complex and merits a work on its own beyond the scope of this paper. We just want to point out that this detailed analysis can be achieved within the general framework presented here.

Finally, we stress that these particular edge-state BICs are unrelated to another kind of `bulk' BICs found in graphene quantum-dot structures~\cite{Gonzalez10} when a particular symmetry is present which induces a destructive interference between two resonant states~\cite{Friedrich85b}.

\subsection{Fano resonance due to chiral symmetry breaking}\label{ssec:t2}

\begin{figure}[tb]
 \centering
 \includegraphics[width=\linewidth]{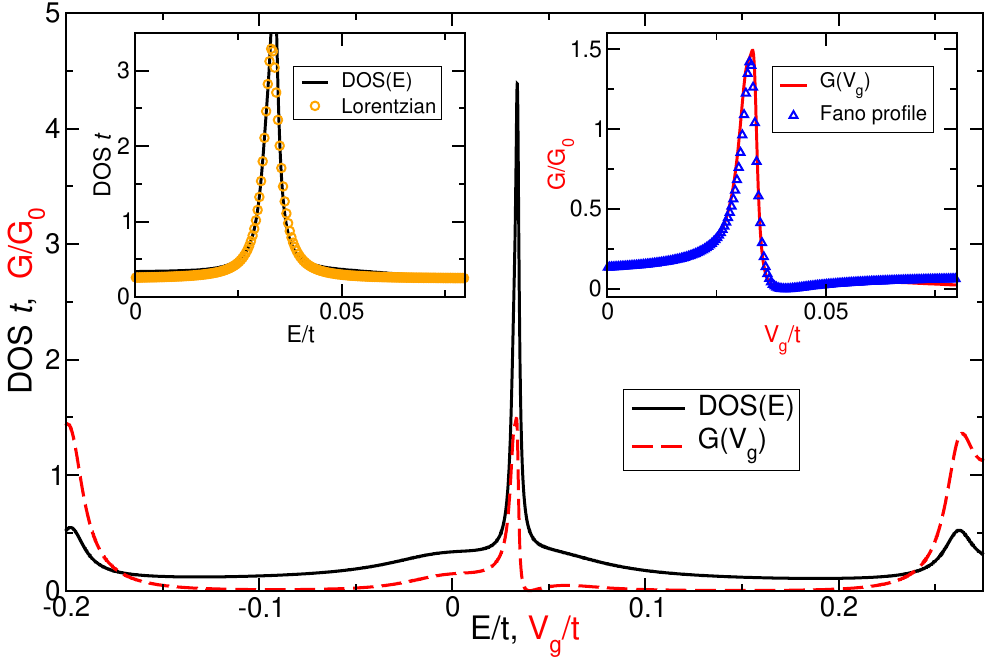}
 \caption{\label{fig:t2}
 (color online)
 DOS (solid black line) and conductance (dashed) for a $10 \times 10$ square lattice connected to four-channel leads, including hopping to next-to-nearest neighbors $t_2=0.01$.
 The insets show the raw data (lines) together with fits (symbols) to a Lorentzian and a Fano profile, respectively.}
\end{figure}

We consider next the effect that a non-negligible next-to-nearest neighbor hopping amplitude, $t_2$, can have on the existence and observability of BICs in lattice systems. To assess this, we consider a system where $\Hch$ contains a small coupling $t_2=0.01$ between sites on the same sublattice. As a consequence, the system is no longer bipartite and zero-energy eigenstates will in general populate both sublattices.
In addition, the small intra-lattice coupling $t_2$ (which leads to a (weak) breaking of the $E \rightarrow -E$ symmetry) will induce a small displacement of all energy levels; in particular states that were at $E=0$ for $t_2=0$ will now have energies $E \sim t_2$.
The DOS and conductance properties of such a system are shown in Fig.~\ref{fig:t2}.
We observe that the DOS peak is now centered around $E_{max} \approx 0.033$, while the conductance shows the characteristic asymmetry of a Fano resonance, peaked at $V_{max}=E_{max}$ and vanishing at $V_{min} \approx 0.039$.
This resonance feature is due to the coupling of the BIC (a ``dark state'' in quantum-optics parlance when $t_2=0$) with the continuum of conduction modes, a coupling originated in the chiral-symmetry breaking perturbation $t_2$.
In Sect.~\ref{sec:disc} we further discuss this resonance in relation with recent experiments in plasmonic nanostructures~\cite{Luky10,Halas11}.

A fit of the DOS to a Lorentzian provides a width for the resonance
$\Gamma \approx 1.7\times10^{-3}$.
Taking this value together with $V_{max}$ and $V_{min}$ allows to estimate the Fano profile (dimensionless) parameter $q$~\cite{Miros10} as negative and of order unity (i.e., $|q|\sim |t| \gg |t_2|$), confirming that continuum and bound state are strongly coupled, resulting in a very asymmetric profile. 


%

\section{Discussion}\label{sec:disc}

\subsection{Insulating behavior around the band center}\label{ssec:insulating}

We have commented the presence of a broad range of energies around $E=0$ for which the system reflects all incoming waves, in case the total number of channels supported by the leads satisfies $\lambda_L+\lambda_R < {\cal N}_\textrm{ch}$.
Moreover, this behavior is not destroyed by a (weak) next-to-nearest neighbor coupling.
The presence of a similar broad insulating band has been predicted to occur in a chain of double quantum dots (QDs) in Ref.~\cite{Gong06}. In that work, the insulating bandwidth was related to the energy bandwidth of an infinite chain of double QDs.
Similarly, Ref.~\cite{Miros05} studied the transmission and reflection of a discrete waveguide with an $N$-state side defect, and observed that the transmission dropped to zero exactly at the energies corresponding to the $N$ eigenstates of the (isolated) defect.
In another study, Voo and Chu~\cite{Voo06} showed the appearance of exponentially-localized states within the continuum of discrete systems attached to low-dimensional leads, and related them to the sharp Fano resonances in the corresponding transport properties.

\begin{figure}
 \centering
 \includegraphics[width=\linewidth]{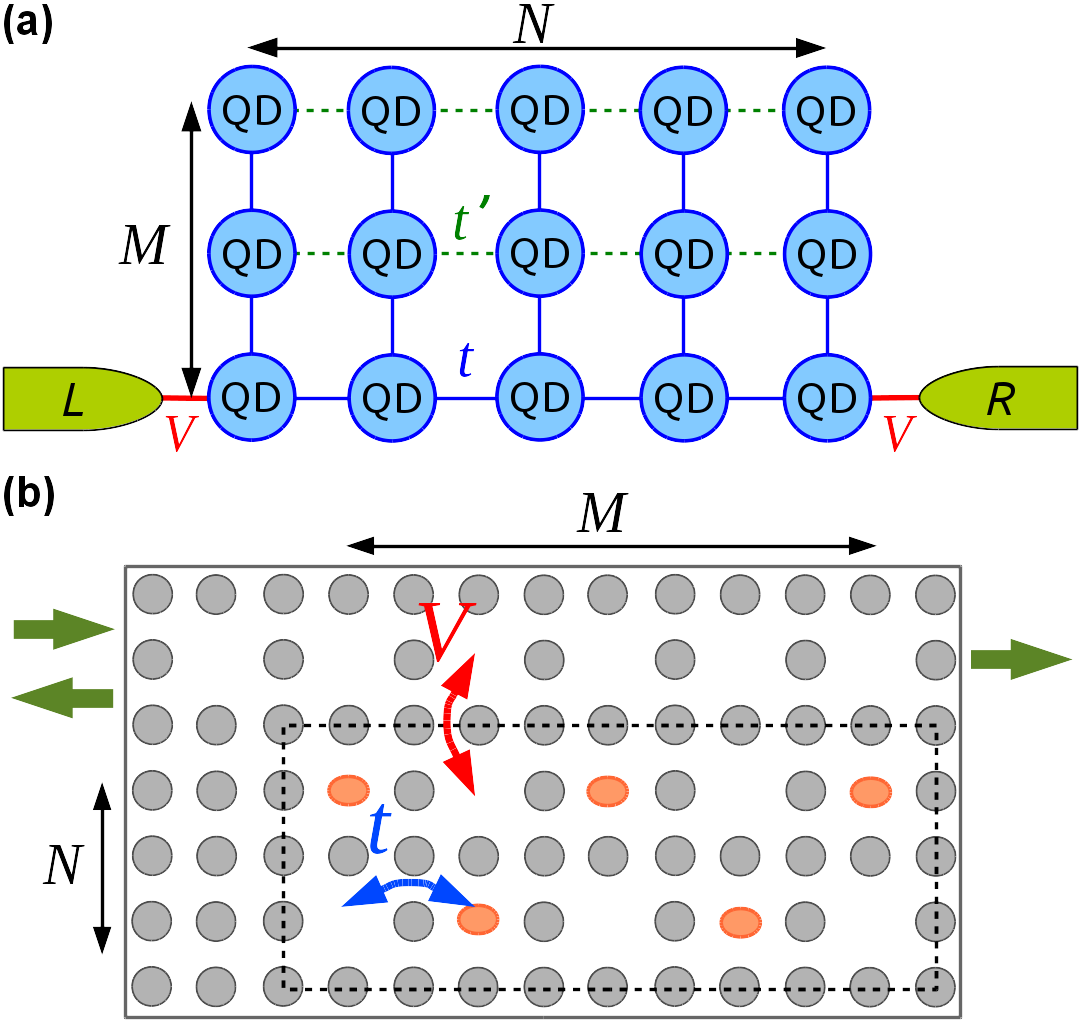}
 \caption{\label{fig:discuss}
 (color online)
 Systems on which the appearance of similar phenomena related to Fano resonances have been discussed in the literature.
 \textbf{(a)} $N\times M$ array of quantum dots with horizontal hopping amplitude $t'$ between QDs on rows above row \#1, coupled to leads $L$ and $R$; for $M=2$ and $t'=0$, this is a chain of $N$ double QDs as in Ref.~\cite{Gong06}.
 \textbf{(b)} Photonic crystal (PC) formed by an array of silica rods (gray circles) in a matrix.
 The row with missing rods at the top forms a waveguide that transmits or reflects light coming to the PC (thick green arrows).
 There is an $N \times M$ array (dashed square) of defects (ellipses, indicating rods of a different refractive index, and additional missing rods) coupled to the waveguide. The $N$-defect chain in Ref.~\cite{Miros05} corresponds to $M=1$.
 }
\end{figure}

Our earlier discussion in Sect.~\ref{sec:chiral} allows us to see these results from a more general perspective.
For instance, a chain of $N$ double QDs, Fig.~\ref{fig:discuss}a, is equivalent to the system in Fig.~\ref{fig:square_lattice}a with $N\times 2$ sites for the case that the hopping between different columns $t'=0$. In this case, the system Hamiltonian, $\Hch$, can support at most ${\cal N}_\textrm{ch}=2$ zero-energy eigenstates. According to our discussion, upon connecting the system to two leads, this brings about the zero conductance for all $N \geq 1$ as seen in Fig.~2 of Ref.~\cite{Gong06}.

By contrast, the $N$-defect chain in a photonic crystal of Ref.~\cite{Miros05} amounts to a $N\times 1$ system in our nomenclature, but connected by only one site to incoming and outgoing modes, see Fig~\ref{fig:discuss}b.
The $N$-defect chain supports one zero-energy eigenstate for odd $N$.
This induces the Fano antiresonance at $E=0$ for odd $N$ in Fig.~1 in Ref.~\cite{Miros05}, in analogy with the case of $\lambda=5$ in Fig.~\ref{fig:dosG} above.
For even $N$, on the other hand, there are no zero-energy eigenstates and, hence, no Fano resonances appear at zero energy.

\subsection{Fano resonance in periodic structures}\label{ssec:FanoRes}

The existence of chiral BICs requires the underlying chiral symmetry that allows to write the system Hamiltonian in the form~\eqref{H:chiral}, and that the coupling to the leads is done through $\lambda_L+\lambda_R < {\cal N}_\textrm{ch}$ modes.
As we saw in Fig.~\ref{fig:dosG}, systems with chiral symmetry with exactly ${\cal N}_\textrm{ch}$ outcoupling modes present a narrow Fano antiresonance at the band center, pointing to perfect destructive interference (vanishing Fano parameter $q=0$) between transmission through the BIC and the continuum of lead modes~\cite{Miros10}.
On the other hand, when the chiral symmetry is broken by the $t_2$ hopping, there is destructive interference between BIC and continuum modes, but this time with $q \neq 0$, so that the conductance has a narrow peak around the new `bare' BIC position, cf. Fig.~\ref{fig:t2}.

In recent years, the development of new fabrication methods of nanostructures, from nanoparticle arrays~\cite{Christ07,Hao08,Yan09ACS,Liu09,Fan10} to photonic crystals~\cite{Hsu13,Bellec13,photonics_book} has lead to a growing interest in the design and production of materials that allow to realize tunable Fano resonances~\cite{Hao08,Fan10,Lassiter10,Zhang12} with a broad range of potential applications, from photon switches to spin filters, see e.g. Refs.~\cite{Miros10,Wu12}.
For example, Ref.~\cite{Hao08} reports that a system composed of a nano\-ring with an inserted nano\-disk can feature a sharp Fano resonance depending on the position of the disk with respect to the ring center. When the disk is centered inside the ring, the coupling between dipolar plasmonic modes of each element leads to a ``superradiant'' and a ``sub-radiant'' collective modes. However, when the disk is off-center, these dipolar modes couple also to the quadrupolar plasmons, which were previously ``dark'', i.e., they behaved as a BIC. As a consequence of this coupling, the previously ``super-radiant'' mode presents now a sharp Fano resonance, pointing to the underlying quadrupolar ``dark'' mode.

In their experiments, Hao \textit{et al.} proved~\cite{Hao08} that it was possible to control the position and shape of the Fano resonance by modifying the overall size or other properties of the system.
In analogy with this ability to control the particular band center and bandwidth transmitted or reflected by such systems~\cite{Christ07,Hao08,Yan09ACS,Liu09,Fan10}, from our results in Sect.~\ref{ssec:t2} we see that one can control the position and shape of the Fano resonance in Fig.~\ref{fig:t2} by changing the value of $t_2$. In experimental setups, this could be realized, e.g., by modifying the properties of the elemental units of the lattice or the distance between them, e.g., the distance between QDs in QD nanocrystals~\cite{QD_book} or between dielectric resonators in a microwave cavity~\cite{Bellec13}, etc. In order to provide a more specific assessment of the requirements of such implementations, we present in the following section a brief discussion of a number of potential experiments to test our results.

\section{Proposals for the Observation of chiral BICs}\label{sec:expt}

The existence of chiral bound states in the continuum as presented in the previous sections is rooted in very general assumptions: that the system accepts a lattice description, and that the lattice is (to a good approximation) bipartite. Many physical systems of interest can be described in these terms and, therefore, can feature chiral BICs. Without trying to be particularly exhaustive we can cite
a wide variety of wave systems (both classical and quantum) in bipartite structures:  electrons traveling through quantum dot arrays~\cite{Gong06,Miros10}, photonic crystals~\cite{Miros05,Bittner12}, nanophotonic structures~\cite{Hao08,Luky10}, exciton-polaritons in square lattices~\cite{Kim11}, etc. Below, we discuss for concreteness two experimental setups that, due to their flexibility of construction and measurement precision, appear as particularly good candidates to observe these phenomena.

\subsection{Microwave dielectric resonators}\label{ssec:mw}

\begin{figure}
 \centering
 \includegraphics[width=\linewidth]{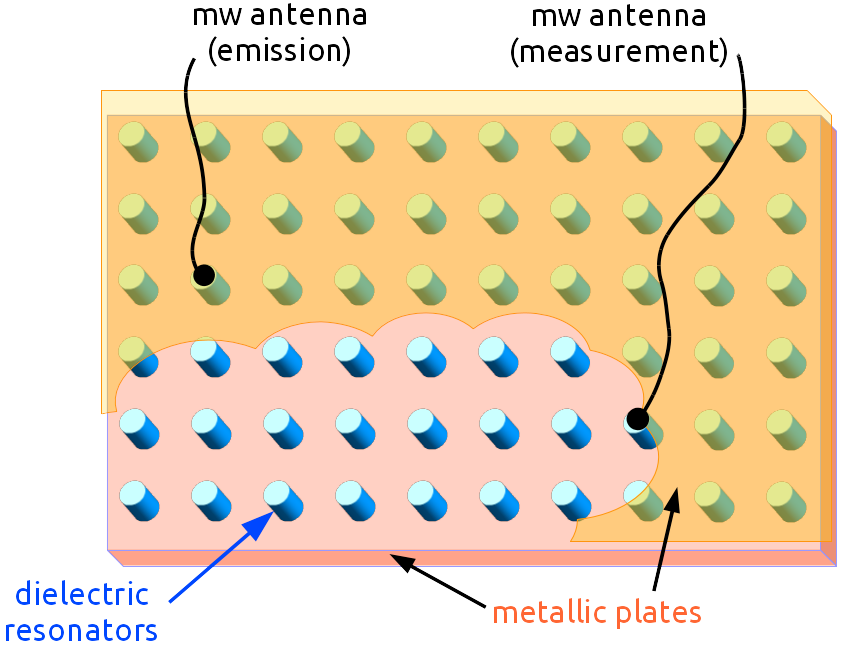}
 \caption{\label{fig:expt-mw}
  (color online)
  \textit{Microwave resonator setup to observe chiral BICs.}
  An array of cylindrical dielectric resonators (blue cylinders) is placed between two parallel metallic plates (top and bottom layers). The conductance and DOS of the system can be probed with movable microwave antennas (black curves).
 }
\end{figure}

A set of identical cylindrical resonators is placed in between two metallic plates constituting an electromagnetic cavity, see Fig~\ref{fig:expt-mw}.
The evanescent field induces a weak coupling between the resonators that can be modelled by a tight-binding Hamiltonian~\cite{Kuhl10}. The resonators can be placed following an arbitrary geometrical arrangement. In experiments by Bellec {\em et al.} they were placed following a square lattice and a honeycomb lattice (artificial graphene)~\cite{Bellec13}. An antenna moving through the system was used to measure the local DOS and the eigenstates of the structure. A combination of antennas could, in principle, be used to measure the transmission properties of such a system. In these experiments next-to-nearest neighbor coupling was not completely suppressed and the density of states was slightly asymmetric with respect to the band center.
We expect that one could then observe the Fano resonances related to the zero-mode states and chiral BICs discussed in Sect.~\ref{ssec:t2}. Changing the value of the next-to-nearest neighbor (NNN) hopping with respect to the nearest-neighbor (NN) hopping terms, which can be done by controlling the spatial distance between the resonators~\cite{Bellec13}, it would be possible to change the position and width of the Fano resonance.

\subsection{Atoms in optical lattices}\label{ssec:optlat}

The group of T. Esslinger at ETH has used cold fermionic atoms to
engineer a cold-atom analogue of electron transport in mesoscopic systems~\cite{Brantut2012,Brantut2013}.
In this setup, a large sample of quantum-degenerate $^6$Li atoms is split into two unequal ``reservoirs'' by means of a blue-detuned laser beam that creates a repulsive potential except for a narrow two-dimensional (2D) channel linking the reservoirs~\cite{Brantut2012,Brantut2013}.
More recently, they have observed quantized conductance through the channel by further constraining it using a mask to realize a one-dimensional (1D) channel with a finite number of transverse modes, whose population can be controlled~\cite{Krinner2014}---a cold-atom analogue of a quantum wire (QW).

\begin{figure}
 \centering
 \includegraphics[width=\linewidth]{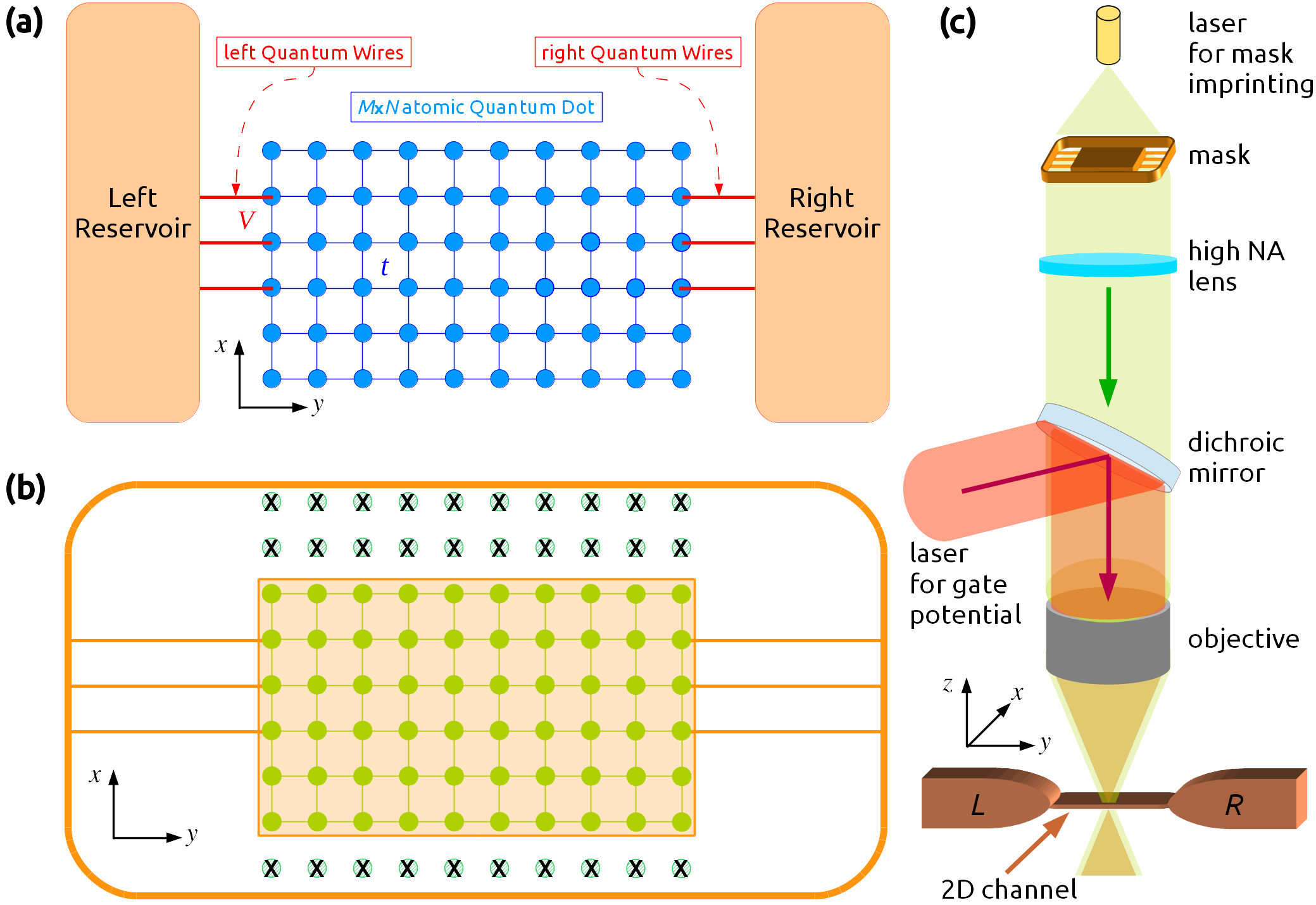}
 \caption{\label{fig:expt-atom}
 (color online)
 \textit{Cold atom setup to observe chiral BICs.}
 \textbf{(a)} Left and right reservoirs of degenerate fermions are connected through a 2D lattice (blue dots). The reservoirs are linked to the lattice by 1D `atomic quantum wires' supporting a small number of transverse channels~\cite{Krinner2014}.
 \textbf{(b)} Mask to imprint the AQWs and a large central ``allowed island'' (orange square).
 Note that lattice sites outside the allowed island would be unreachable (crossed circles) due to the blue-detuned laser beam (not pictured) that creates the 2D channel.
 \textbf{(c)}
 Experimental setup scheme, following~\cite{Krinner2014}.
 }
\end{figure}

An extension of this scheme, sketched in Fig.~\ref{fig:expt-atom}a, would enable testing our predictions.
Here, each atomic reservoir is connected to a finite number, $\lambda$, of such atomic quantum wires (AQWs), while these are connected to a central bipartite lattice where the atoms can hop from site to site.
The AQWs can be realized by imprinting a mask in a manner analogous to Ref.~\cite{Krinner2014}, while the central lattice could be realized in two ways. A first option would consist of imprinting a large ``allowed region'' between the AQWs, on top of which one would project a 2D optical lattice, in a way similar to~\cite{bakr2009,sherson2010}, see Fig.~\ref{fig:expt-atom}b.
Alternatively, one could create a mask containing the structure for both the AQWs and the lattice.

This experimental approach enables to control the NN and NNN hopping amplitudes $t,t_2$ by means of the lattice spacing, while the coupling and number of leads is given by the design of the mask as well as a gate voltage that controls the number of open transverse channels on each AQW~\cite{Krinner2014}.
The conductance of the lattice is then measured by creating a difference in the chemical potential of the two reservoirs, and measuring the relative particle number after a finite time~\cite{Brantut2012,Brantut2013,Krinner2014}.

\section{Conclusions}\label{sec:conc}

To summarize, we have explored the appearance of bound states in the middle of the band of chiral quantum systems connected to leads defining a continuum of scattering states. The number of these bound states in the continuum (BICs) depends on the rank of the matrix that couples the two subsystems defining the chiral symmetry and on the number of states in each subsystem directly connected to the scattering states in the leads. A very general example of the chiral systems we have considered is a system defined by a discrete bipartite lattice with hopping matrix elements only between sites in different sublattices and connected to leads only through a small number of sites. These lattice Hamiltonians can model very different physical systems where we expect observable consequences of the presence of BICs: quantum dot arrays, atoms in optical lattices, microwave resonators, etc. We have explored the consequences in the transport properties of such systems and showed that they feature zero two-terminal conductance at a value of the energy where the density of states diverges. In limiting cases, when the number of coupling channels is equal to the number of zero-energy states or when a small perturbation weakly breaks the chiral symmetry, the conductance null transforms into a Fano resonance.

The theory presented in this work provides a general framework for understanding the presence of BICs and Fano resonances in finite lattices with chiral symmetry. We have discussed how the present framework allows to understand in a general way many results found in the literature concerning Fano resonances and BICs in specific configurations of quantum dot arrays, photonic lattices, or even dark states in transport configurations~\cite{Gong06}.
Finally, we have proposed two experimental realizations where it should be possible to test our predictions experimentally: microwave dielectric resonators and atoms in optical lattices.
We expect that these results will allow further progress in quantum transport studies in lattice structures and in the design of tunable quantum devices~\cite{Miros10,Hao08,Wu12}.

\acknowledgments
We acknowledge fruitful discussions with V. Fern\'andez-Hurtado, J. J. Garc{\'\i}a-Ripoll and L. Tarruell.
This work was supported by Spanish MINECO projects No.~FIS2012-33022 and FIS2012-34479, ESF Programme POLATOM, and the JAE-Doc program (CSIC
and European Social Fund).

\bibliography{biblio-bic}

\end{document}